\documentstyle[aps,preprint]{revtex}

\tightenlines

\begin{document}

\title{Subbarrier heavy ion fusion enhanced by nucleon transfer
and subbarrier fusion of nuclei far from the line of
$\beta$-stability}

\author{V.Yu. Denisov}

\address{Institute for Nuclear Research, Prospect Nauki 47, 252028
Kiev, Ukraine \footnote{E-mail: denisov@kinr.kiev.ua, Permanent
address}\\ and \\ GSI-Darmstadt, Planckstrasse 1,
D-64220 Darmstadt, Germany \footnote{E-mail: denisov@gsi.de}}

\maketitle

\begin{abstract}
We discuss a model for the description of subbarrier fusion of
heavy ions which takes into account the coupling to the low-energy
surface vibrational states and to the few-nucleon transfer with
arbitrary reaction $Q$-value. The fusion reactions
$^{28,30}$Si+$^{58,62,64}$Ni, $^{40}$Ca+$^{90,96}$Zr,
$^{28}$S+$^{94,100}$Mo, $^{16,18,20,22,24}$O+$^{58}$Ni and
$^{28}$Si+$^{124,126,128,130,132}$Sn are analyzed in detail. The
model describes rather well the experimental fusion cross section
and mean angular momentum for reactions between nuclei near the
$\beta$-stability line. It is shown that these quantities are
significantly enhanced by few-nucleon transfer with large positive
$Q$-value. A shape independent parameterization of the heavy ion
potential at distances smaller then the touching point is
proposed. \end{abstract}

PACS number(s):
25.60.Je, 
25.60.Pj, 
25.70.Hi, 
25.70.Jj 


\section{Introduction}

Heavy ion fusion reactions at energies below or near the Coulomb
barrier have received considerable attention [1-33]. Recently
many different mechanisms were discussed for the description of
subbarrier fusion reactions as coupling to the low-energy excited
states, nucleons transfer, deformation of ions or neck formation
during barrier penetration [3-33]. Fusion cross sections are
strongly enhanced at energies below barrier by the coupling to
both the low-energy surface vibrational states [3-14,27-31] and
the few-nucleon transfer channels [3-11,15-23,30,32]. Models which
take into account the coupling to the low-energy surface
vibrational states as well as to the few-nucleon transfer channels
describe well the fusion cross section $\sigma_{\rm fus}(E)$ and
mean angular momentum of compound nucleus $<L(E)>$ for reactions
between nuclei near the line of $\beta$-stability.

The coupling potentials between the ground state channel and
channels connected with the low-energy vibrational states are
well-known [1,3-14]. Therefore the theory of fusion cross section
enhancement due to coupling to low-energy vibrational states is
well developed. In contrast to this, the coupling potential for
transfer channels as a rule is not known with good accuracy. This
potential may be fixed by studying quasi-elastic transfer
reactions. Unfortunately, experimental information on quasi-elastic
transfer reactions often is not available. Therefore the
description of fusion cross section enhancement due to transfer
reactions is based on the fitting parameters. Moreover, the radial
shape of the coupling potential for transfer channels has various
forms in the different models. For example, the radial dependence
of the transfer coupling potential is chosen in the forms $F
\exp[\alpha (r - R_{12})]$ [3,14] or $F \exp[\alpha (r -
R_{12})]/r$ \cite{corradi93}, where $F$ is a coupling constant,
$\alpha$ is a constant related to the separation energy of
transferred particles or is taken from systematics \cite{ccfusb},
$r$ is the distance between the centre of mass of the ions,
$R_{12}=R_1+R_2$, $R_1$ and $R_2$ are the radii of the ions.
Sometimes the transfer coupling potential has a similar form as the
coupling potential to the low-energy vibrational states [11,19,29].
The value of the transfer constant is not fixed a priori and
sometimes is chosen by the fitting experimental data
\cite{expSPd}.

The simplified coupled-channel code CCFUS \cite{ccfus} is often
used for the analysis of subbarrier fusion of heavy ions
[3,7-9,19,28-31]. As pointed out by Landowne (see Refs.
\cite{vandenbosch,expSMo2}) the CCFUS model overestimates the
contributions of the transfer channels in the case of large
positive $Q$-value of the transfer reaction.

Recently, by using radioactive ion beams an experimental
possibility was available to study fusion reactions between a
nucleus far from the line of $\beta$-stability \cite{expAlAu}. The
fusion reactions induced by such colliding systems may be strongly
enhanced by transfer reactions with a large positive $Q$-value.
Therefore it is necessary to develop a simple model describing
fusion reactions that takes into account the coupling both to the
low-energy surface vibrational states and to the transfer channels
with their large positive $Q$-values.

The direct solution of coupled reaction channel equations is a
difficult numerical problem, see for example \cite{idm}. It is
shown in Ref. \cite{stefanini} that the heavy ion fusion cross
sections calculated with the "exact" microscopic code is in good
agreement with the one obtained by using the CCFUS model
\cite{ccfus}, when transfer channels are not important. Therefore
we consider the coupling to the low-energy surface oscillations
during the fusion process in the same manner as in the CCFUS
model. We treat the coupling to the transfer channels in the DWBA
approximation \cite{bass}, which describes well the quasi-elastic
transfer reactions near barrier \cite{rehm}. The nucleon transfer
during the fusion reaction will be considered in the WKB
approximation \cite{ll3}.

The probability of barrier penetration is determined by the
action in the WKB approximation \cite{ll3}. We consider that
nucleon transfer takes place during barrier penetration. Therefore
the action splits into three terms according to the Landau method
of complex classical paths for transitions in systems with
arbitrary degrees of freedom, see for details \S 52 in \cite{ll3}.
The tunneling from the outer turning point to the distance
$r_{tr}$, where the transfer of a particle takes place, is
described by the first term. The second term relates to the
probability of the nucleon transfer process at the distance
$r_{tr}$ between the ions. The third term describes the tunneling
from $r_{tr}$ to the inner turning point. The enhancement of the
subbarrier fusion reaction due to the nucleon transfer process has
not been considered in such approximation.

We describe the probability of nucleon transfer by using the
semiclassical model \cite{bass} developed for the description of
subbarrier nucleon transfer between two ions. The model discussed
in \cite{bass} does not employ a transfer coupling constant.
Therefore in our model it is not necessary to know from other
experimental data the value of the coupling constant related to
transfer processes for the calculation of the subbarrier fusion
cross section. In contrast to previous considerations our method
is valid for arbitrary $Q$-values of transfer the reaction.

Our model is discussed in detail in the Sec. 2. In Sec. 3
$\sigma_{\rm fus}(E)$ and $<L(E)>$ are calculated within the
proposed model and compared with experimental data for fusion
reactions induced by nuclei located along the $\beta$-stability
line. The fusion reaction cross section and mean angular momentum
of the compound nuclei obtained in this model are analyzed in the
case of fusion reactions between $\beta$-stable nuclei and nuclei
near the neutron drip line in Sec. 4. Summary and conclusions are
presented in Sec. 5.

\section{Subbarrier heavy ion fusion enhanced by nucleon transfer}

The system of coupled channel equations in the case of coupling to
the low-energy vibrational states has the form [1,3-5,10,11,13,14]
\begin{equation} \left[ -\frac{\hbar^2}{2\mu_i} \frac{d^2}{dr^2} +
\frac{\hbar^2 \ell_i (\ell_i+1) }{2\mu_i r^2} + V(r) - Q_i - E
\right] \varphi_i(r) = - \sum_j V_{ij}(r) \varphi_j(r),
\end{equation} where $\psi_i(r) = \varphi_i(r)/r$ is the wave
function, $\mu_i$ is the reduced mass, $\ell_i$ is the value of
the orbital momentum in units of $\hbar$, $V(r)$ is the ion-ion
interaction potential, $Q_i$ is the $Q$-value of the reaction in
channel $i$, $E$ is the collision energy and $V_{ij}(r)$ is the
coupling potential. The coupling potential between the ground
state and the channels connected with the low-energy surface
vibrational state of multipolarity $\lambda$ is given by
[1,3-5,10,11,13] \begin{equation} V_{0i} = \frac{ \beta_i
R_i}{\sqrt{4\pi}} \left[ \frac{d V_{\rm i-i}(r)}{d r} +
\frac{3}{2\lambda+1} \frac{z_1 z_2 e^2
R_i^{\lambda-1}}{r^{\lambda+1}} \right]. \end{equation} Here
$V_{\rm i-i}(r)$ is the nuclear part of the ion-ion interactions
$V(r)$, $z_1$ and $z_2$ are the proton numbers, $e$ is the proton
charge and $\beta_i R_i$ is the deformation length of the $i$-th
vibrational state in the nucleus with radius $R_i$.

As in [3,4,10,11,13,14] we propose that all reduced masses $\mu_i$
and orbital momenta $\ell_i$ are equal in all channels related to
vibrational excitations. Then by taking the radial dependence of
the coupling potential at the barrier position $V_{ij}(r) =
V_{ij}({\overline R})$ we diagonalize the system (1) with the help
of the substitution \begin{equation} \varphi_i(r) = \sum_k U_{ik}
\xi_k (r), \end{equation} where $U_{ik}$ is the transformation
matrix and $\xi_k(r)$ is the wave function
(eigenvector). The coupling matrix ${\cal M}_{ij}$ takes the form
\begin{equation}\sum_{ij} U_{ki} {\cal M}_{ij} U_{jl} = \sum_{ij}
U_{ki} [-Q_i \delta_{ij} + V_{ij}({\overline R})] U_{jl}
=\epsilon_k \delta_{kl} \end{equation} and upon diagonalization we
find the eigenvalue $\epsilon_k$. In this case the partial
fusion cross section $\sigma(E, \ell)$ is equal to
[3,5,10,11,13-14] \begin{equation} \sigma(E, \ell) = \frac{\pi
\hbar^2}{2\mu E} (2\ell+1) \sum_k |U_{k0}|^2 T(E, {\cal V}_{\ell
k}), \end{equation} where $T(E, {\cal V}_{\ell k})$ is the
transmission coefficient obtained for the one-dimensional
effective potential ${\cal V}_{\ell k}$ \begin{equation} {\cal
V}_{\ell k}(r)=V_\ell(r)+\epsilon_k = V(r) + \hbar^2
\ell(\ell+1)/(2\mu r^2) + \epsilon_k. \end{equation} We conclude
from (5) that the partial cross section for fixed $E$ and $\ell$
is determined by the sum of the transmission coefficients $T(E,
{\cal V}_{\ell k})$ obtained for the effective potential ${\cal
V}_{\ell k}$ with the weights $|U_{k0}|^2$. The effect of fusion
cross section enhancement due to the coupling to the low-energy
vibrational states is related to the smallest eigenvalue
$\epsilon_k$, which is negative.

The total fusion cross section is equal to
\begin{equation} \sigma_{\rm fus}(E) = \sum_\ell \sigma(E,\ell).
\end{equation}

Let us consider the transfer reaction in the DWBA approach,
which described well nucleon transfer reactions near and below
barrier \cite{bass}. In the DWBA approximation we neglect the
influence of the transfer channels on the channels without
transfer and on other transfer channels. In this case the matrix
${\cal M}$ has a box structure. Each box of the matrix ${\cal M}$
in (4) is similar to the respective box without transfer. For
each transfer channel we have an enhancement described by Eqs.
(4)-(6). For the sake of simplicity we propose that the values of
$\epsilon_k$ and $|U_{k0}|^2$ for each specific transfer channel
do not differ much from the ones obtained in (4) without transfer.
Our proposal is based on a small variation of both the energies
and the deformation lengths of the vibrational states in heavy
nuclei which differ by several nucleons. In this case the partial
fusion cross section in the transfer channel $f$ is determined
also by Eqs. (4)-(6), but the transmission coefficient should be
calculated by taking into account the few-nucleon transfer.

If the energy of collision is smaller than the barriers of the
effective potentials before and after nucleon transfer and if the
transfer occurred at the distance $r_{tr}$, then the transmission
coefficient may be written as \begin{equation} T(E,{\cal V}_{\ell
k}^i,{\cal V}_{\ell k}^f ) = 1/\{1+\exp[{\cal A}(E,{\cal V}_{\ell
k}^i,{\cal V}_{\ell k}^f,r_{tr}) ]\}, \end{equation} where the
action ${\cal A}(E,{\cal V}_{\ell k}^i,{\cal V}_{\ell
k}^f,r_{tr})$ is given by \begin{equation} {\cal A}(E,{\cal
V}_{\ell k}^i,{\cal V}_{\ell k}^f,r_{tr}) = {\cal A}^i(E,{\cal
V}^i_{\ell k},r_{tr}) + {\cal A}^{tr}(E,r_{tr}) + {\cal
A}^f(E,{\cal V}^f_{\ell k},r_{tr}). \end{equation} Here we apply
the Landau method of complex classical paths for transitions in
systems with arbitrary degrees of freedom, see for details Eq.
(52.1) and related text in \S 52 in \cite{ll3}. The action
\begin{equation} {\cal A}^i(E,{\cal V}^i_{\ell k},r_{tr}) =
(2/\hbar) \int^{r_{\ell k}^i}_{r_{tr}} \sqrt{2 \mu_i(r) ({\cal
V}^i_{\ell k}(r)-E) } dr , \end{equation} describes the tunneling
of ions in an effective potential before nucleon transfer ${\cal
V}^i_{\ell k}$ from the outer turning point $r_{\ell k}^i$ up to
$r_{tr}$, the action ${\cal A}^f(E,{\cal V}^f_{\ell k},r_{tr})$
\begin{equation} {\cal A}^f(E,{\cal V}^f_{\ell k},r_{tr}) =
(2/\hbar) \int_{r_{\ell k}^f}^{r_{tr}} \sqrt{2 \mu_f(r) ({\cal
V}^f_{\ell k}(r)-E) } dr \end{equation}is related to the tunneling
of ions in an effective potential after nucleon transfer ${\cal
V}^f_{\ell k}$ , \begin{equation} {\cal V}^f_{\ell k} (r) =
V^f_{\ell} (r) + \epsilon_k - Q_{\rm transfer}^f \end{equation}
from the point $r_{tr}$ to the inner turning point $r_{\ell k}^f$
of the effective potential ${\cal V}^f_{\ell k} (r)$. Here $Q_{\rm
transfer}^f$ is the $Q$-value of the transfer reaction in the
channel $f$.

In the case of $m$-neutron transfer during barrier penetration in
fusion of heavy ions the action ${\cal A}^{tr}(E,r_{tr})$
connected with the nucleon transfer process may be written as
\begin{equation} {\cal A}^{tr}(E,r_{tr}) = (2/\hbar) \sum_{i=1}^m
\sqrt{2 M {\cal E}_i} ( r_{tr} - R_{12} - \delta ). \end{equation}
This form of the action describes the tunneling of $m$ neutrons
between spherical square potential wells of the colliding ions. In
(13) we introduced a parameter $\delta$ because due to the finite
diffuseness of the realistic nucleon-nucleus potential the barrier
for the transferred nucleon disappears at the finite distance
$\delta>0$ between the surfaces of the ions. The action (13) is
often used for the description of subbarrier neutron transfer
reactions between heavy ions [1,32,35-39].

The wave function of the transferred nucleon may concentrate more
in the volume or in the surface part of the nucleus. Due to this
the nucleon transfer amplitude related to the overlap integral of
the wave functions can have its maximum of transfer probability at
relatively larger or smaller distances between colliding ions. It
is possible to take into account this fine effect by a small
variation of the parameter $\delta$ in (13).

The distance $r_{tr}$ at which the nucleon transfer takes place is
determined from the principle of minimal action, see \S 52 in
\cite{ll3}. The trajectory of tunneling obtained by taking into
account the few-nucleon transfer between heavy ions has its minimum
value of the action (9) and its maximum value of the transmission
coefficient (8). The few-nucleon transfer is especially important
when $Q^f_{\rm transfer} \gg 1$ MeV and the action (11) is small.

The action ${\cal A}(E, {\cal V}_{\ell k}^i,{\cal V}_{\ell k}^f,
r_{tr})$ is a function of the $Q$-value of the transfer reaction
and of the separation energies ${\cal E}_i$ of the transferred
nucleons. Therefore the most favorable condition for the
enhancement of subbarrier fusion due to few-nucleon transfer takes
place at small separation energies of the transferred nucleons
${\cal E}_i$ and at large positive $Q$-values of the transfer
reactions.

The expression (8) for the transmission coefficient is valid for
collision energies $E$ smaller than the effective barriers
${\overline{\cal V}}^i_{\ell k}$, before and ${\overline{\cal
V}}^f_{\ell k}$, after the few-nucleon transfer. In the case
${\overline{\cal V}}^f_{\ell k} < E < {\overline{\cal V}}^i_{\ell
k}$ and $r_{tr} > {\overline R}^f_{\ell k}$ the transmission
coefficient has the form \begin{equation} T(E,{\cal V}^i_{\ell
k},{\cal V}^f_{\ell k}) = 1/\{1+\exp[ {\cal A}^i(E,{\cal
V}^i_{\ell k},r_{tr}) + {\cal A}^{tr}(E,r_{tr}) ]\} \; \; T_{\rm
HW}(E,{\cal V}^f_{\ell k}). \end{equation} Here ${\overline
R}^f_{\ell k}$ is the barrier distance of the effective potential
${\cal V}^f_{\ell k}$, $T_{\rm HW}(E,{\cal V}^f_{\ell k})$ is the
transmission coefficient of the effective barrier after transfer
obtained in the Hill-Wheeler approximation \cite{h-w} and taking
into account the reflection during barrier penetration. The
subbarrier tunneling of ions before the nucleon transfer and the
subbarrier nucleon transfer are described by the first factor in
(14). The second factor in (14) is related to the transmission
above the barrier between the ions after nucleon transfer.

If ${\overline{\cal V}}^f_{\ell k} < E < {\overline{\cal
V}}^i_{\ell k}$ and $r_{tr} < {\overline R}^f_{\ell k}$, then we
should take into account the decay of the system after the
few-nucleon transfer. In this case the transmission coefficient
may be written as \begin{equation} T(E,{\cal V}^i_{\ell k},{\cal
V}^f_{\ell k}) = 1/\{1+\exp[ {\cal A}^i(E,{\cal V}^i_{\ell
k},r_{tr}) + {\cal A}^{tr}(E,r_{tr}) ]\} (1-T_{\rm HW}(E,{\cal
V}^f_{\ell k})). \end{equation}

We use the transmission coefficient in the Hill-Wheeler
approximation at the high collision energy $E > {\overline{\cal
V}}^f_{\ell k} $ and $E > {\overline{\cal V}}^i_{\ell k}$ and do
not employ the enhancement of fusion due to nucleon transfer. The
expressions (14) and (15) are written for the case $Q_{\rm
transfer} > 0$ and may easily be transformed to the case $Q_{\rm
transfer} < 0$.

The compound nucleus is formed in any transfer channel. Therefore
the total cross section is the sum of (5) and of all possible
transfer channels $f$, i.e. \begin{equation} \sigma_{\rm fus}(E) =
\frac{\pi \hbar^2}{2\mu E} \sum_\ell (2\ell+1) \sum_k |U_{k0}|^2
[T(E, {\cal V}^i_{\ell k})+ \sum_f T(E,{\cal V}^i_{\ell k},{\cal
V}^f_{\ell k})]. \end{equation} Note that the contributions of the
channels with $Q_{\rm transfer} \approx 0$ to the total cross
section are small and negligible for $Q_{\rm transfer}<<1$ MeV due
to the exponential dependence of the transmission coefficient in
the actions. Here we are not consider special cases when the
transferred particle(s) exchanged between identical nuclei as in
the cases of $^{12}$C+$^{13}$C \cite{idm} or $^{58}$Ni+$^{60}$Ni.

Now we determine the interaction potential between two ions at
distance $r$, \begin{equation} V(r) = z_1 z_2 e^2/r + V_{\rm i-i}(r).
\end{equation} Many different parameterizations of the nuclear
interaction potential $V_{\rm i-i}(r)$ between spherical ions
[1-7,39] are available. We choose the Krappe-Nix-Sierk $V_{\rm
KNS}(r)$ \cite{KNS} potential in our calculation for $r \geq
R_{12}=R_1+R_2$. The potential $V_{\rm KNS}(r)$ and the Coulomb
energy depend on the shape of the ions at $r< R_{12}$. We would
like to avoid a shape dependence of the potential $V(r)$. Hence we
use a parameterization of the interaction potential $V(r)$ for $r <
R_{12}$ in the form \begin{equation} V_{\rm fus}(r)= -Q_{\rm fus}
+ x^2 (c_1+c_2 x), \end{equation} where $Q_{\rm fus}$ is the
$Q$-value of the fusion reaction obtained by using the mass table
\cite{wabstra} or by using the mass formula \cite{weizs}, $x=
(r-R_{\rm fus}) / (R_{12}-R_{\rm fus})$, $R_{\rm fus}$ is the
distance between the centers of mass of the left and right parts
of the spherical compound nuclei. The coefficients $c_1$ and $c_2$
are obtained by matching at the touching point $R_{12}=R_1+R_2$
for the potentials $V(r)$ (17) and $V_{\rm fus}(r)$ (18) and for
its derivatives. We propose a quadratic dependence of $V_{\rm
fus}(r)$ at the point $x=0$ because the potential (deformation)
energy of the highly excited compound nucleus is minimum for the
spherical shape, i.e. at $x=0$.

The reduced mass $\mu$ for $r>R_{12}$ is determined by using a
standard expression, see for example \cite{bass}. The reduced mass
in (10) and (11) for $r<R_{12}$ is a function of $r$. We used the
parameterization of $\mu (r)$ introduced in \cite{masspar}
\begin{equation} \mu_{i(f)}(r) = \mu_{i(f)} \{ (17/15) \; k
[(R_{12}-r)/ (R_{12}-R_{\rm fus})]^2 \exp[-(32/17) \; (r/R_{\rm
fus}-1)] + 1 \} , \end{equation} where $k=16$ \cite{masspar}. This
semi-empirical dependence of the reduced mass is successfully used
in the calculation of the lifetime of heavy nuclei for fission
\cite{masspar} and cluster \cite{cluster} decays.

Note that if we neglect the influence of the transfer channels
then the treatment of enhancement of coupled channels due to the
low-energy excitations in our model is similar to the CCFUS model
\cite{ccfus}. In this case the difference between our model and
the CCFUS model is related to the calculation of the transmission
coefficient below barrier. Below barrier this coefficient is
estimated in the CCFUS model by using the Hill-Wheeler
approximation \cite{h-w} but we use the WKB approximation instead
and obtain this coefficient by using the action. (Here we
neglected the difference related to the parameterization of the
nuclear part of the ion-ion potential because the calculations in
both models can be done for the same parameterization of the
nuclear potential.) Hence, if we neglect transfer channels our and
the CCFUS models lead to similar results.

\section{Entrance channel effects at fusion reactions}

Let us consider several fusion reactions between nuclei located
near the line of $\beta$-stability.

First we study isotopic effects in the fusion reactions
$^{28,30}$Si+$^{58,62,64}$Ni. The fusion cross sections calculated
in different approaches for these reactions are compared with the
experimental data \cite{expSiNi} in Fig. 1. The one-dimensional
tunneling model strongly underestimates the experimental fusion
cross sections for the reactions $^{28,30}$Si+$^{58,62,64}$Ni, see
Fig. 1. We obtain similar results if neutron transfer channels
with positive $Q$-value are taken into account, see Fig. 1. We
describe well the experimental fusion cross sections for these
reactions when the coupling to the low-energy $2^+$ and $3^-$
surface excitation states is taken into account. However, we obtain
better agreement with the experimental data for the reactions
$^{28}$Si+$^{62,64}$Ni and $^{30}$Si+$^{58}$Ni when the coupling
to the low-energy vibrational states and to the neutron transfer
channels is taken into account simultaneously, see Fig. 1.

In our calculations we are taking into account 1-, 2-, 3- and
4-neutron transfer channels with positive $Q$-values. The
$Q$-values of transfer reactions are obtained by using the mass
table \cite{wabstra}. The $Q$-values of neutron transfer reactions
for reactions $^{28}$Si+$^{62,64}$Ni and $^{30}$Si+$^{58}$Ni are
small. Here and below we neglect transfer channels with negative
$Q$-value, because the influence of these channels is negligible.
The energies and the deformation parameters of $2^+$ and $3^-$
vibrational states were taken from other experimental data, see in
\cite{expSiNi,lowstate}. These parameters are listed in Table 1.
Here and below for the sake of fitting the experimental fusion
cross section at high energies for these reactions we slightly
change the parameter of the nuclear radii $r_0$ ($R_i=r_0
A_i^{1/3}$) in the KNS potential \cite{KNS}. The values of $r_0$
used in our calculations are also given in Table 1. The values of
$r_0$ for the Si and Ni in Table 1 insignificantly differ from
$r_0=1.18$ fm recommended in \cite{KNS}. We have done the
calculation of the action (13) for $\delta = 0.7$ fm. This value
of $\delta$ is reasonable, because it should be close to the value
of the diffuseness of the realistic nucleon-nucleus potential.

The nuclei $^{62,64}$Ni and $^{30}$Si are donors of neutrons in the
reactions $^{28}$Si+$^{62,64}$Ni and $^{30}$Si+$^{58}$Ni. The
fusion cross sections in collision of $^{28}$Si with Nickel
isotopes are enhanced with increasing the number of neutrons in
Ni. Note that the same compound nucleus is formed in the fusion
reactions $^{28}$Si+$^{64}$Ni and $^{30}$Si+$^{62}$Ni, but the
fusion cross section for the former reaction is larger than for the
latter due to the different values of the parameters of the $2^+$
and $3^-$ surface vibrational states (see in Table 1) and the
various contributions of the transfer channels.

The quality of description of the experimental fusion cross
sections for the reactions Si+Ni in Fig. 1 in our model is similar
to the one obtained by using the CCFUS model in \cite{expSiNi}.
This means that both models lead to similar results for transfer
with the small $Q$-values.

Now we consider fusion reactions with large $Q$-values in the
neutron transfer channels.

The fusion cross sections of $^{40}$Ca+$^{90,96}$Zr \cite{expCaZr}
have been measured recently. The $Q$-values of 2- and 4-neutron
transfer from $^{96}$Zr to $^{40}$Ca are equal to 5.526 MeV and
9.637 MeV respectively. In contrast to this the reaction
$^{40}$Ca+$^{90}$Zr has negative $Q$-values for the transfer of
neutrons.

The Coulomb field at large distance for these reactions is the
same. The heights of barriers have similar values for these
reactions, see Fig. 2. Therefore we may expect that the subbarrier
fusion cross sections for these reactions also are similar.
However, the experimental subbarrier fusion cross sections for the
reaction $^{40}$Ca+$^{96}$Zr is much larger than for the reaction
$^{40}$Ca+$^{90}$Zr, see Fig. 3 and \cite{expCaZr}.

At the beginning we try to describe these reactions by using the
1-dimensional WKB approach. The calculations of $\sigma_{\rm
fus}(E)$ for both reactions in the one-dimensional tunneling
approach yield similar values of the fusion cross sections. But we
can see from Fig. 3 that these calculations strongly underestimate
the experimental data.

The comparison between the experimental data and the theoretical
curves in Fig. 3 is drastically improved for the reaction
$^{40}$Ca+$^{90}$Zr when the low-energy surface vibrational $2^+$
and $3^-$ states in colliding nuclei are taken into account. The
deformations $\beta_{2,3}$ are taken from another experimental
data \cite{lowstate} and are given in Table 1. Nevertheless, the
theoretical curves obtained in this approach for reaction
$^{40}$Ca+$^{96}$Zr still underestimate the experimental data in
Fig. 3.

We performed calculation by taking into account four transfer
channels related to 1-, 2-, 3- and 4-neutron transfer from
$^{96}$Zr to $^{40}$Ca. The heights of barriers of the effective
potentials related to 2-, 3- and 4-neutron transfer channels are
essentially lower then barrier of potential without neutron
transfer, see Fig. 2. Therefore we have significant enhancement of
the subbarrier fusion cross section by taking into account 2-, 3-
and 4-neutron transfer channels. The curves in Fig. 3 associated
to this calculation also underestimate the experimental data. Note
that the enhancement of subbarrier fusion cross section due to
neutron transfer for reaction $^{40}$Ca+$^{96}$Zr in Fig. 3 is
much larger than the ones for the reactions Si+Ni in Fig. 1
because of different $Q$-values of the neutron transfer reactions
for this systems. The model describes the experimental data for
the reaction $^{40}$Ca+$^{96}$Zr when we take into account the
coupling both to the low-energy surface $2^+$ and $3^-$
vibrational states and to the 4 transfer channels, see Fig.  3.
This calculation slightly underestimates the experimental fusion
cross section $^{40}$Ca+$^{96}$Zr for very low energies.  Probably
for this reaction it is necessary to take into account the
coupling to a larger number of excited surface vibrational states,
as done in \cite{expCaZr} for reaction $^{40}$Ca+$^{90}$Zr.

The reactions $^{40}$Ca+$^{90,96}$Zr are also analyzed in
\cite{expCaZr} by taking into account the coupling to the 1- and
 2-phonon surface vibrational states. The theoretical curves
obtained in \cite{expCaZr} describe well the experimental data for
the $^{40}$Ca+$^{90}$Zr and strongly underestimate the
experimental data below barrier for the reaction
$^{40}$Ca+$^{96}$Zr.

Now we consider the fusion reactions $^{28}$Si+$^{94,100}$Mo.
We take into account both the low-energy $2^+$ and $3^-$
surface vibrations and the 1-, 2-, 3-, 4-, 5- and 6-neutron
transfer channels with positive $Q$-values in the calculations of
the fusion cross sections for $^{28}$Si+$^{100}$Mo. For reaction
$^{28}$Si+$^{94}$Mo the coupling to the $2^+$ and $3^-$ surface
vibrations and to the 1- and 2-neutron transfer channels with
positive $Q$-values is taken into account in the calculations.
The deformations $\beta_{2,3}$ are taken from another experimental
data \cite{lowstate} and are listed in Table 1. We can see in Figs.
4 that our calculations describe well the experimental fusion
cross sections for the reactions $^{28}$Si+$^{94,100}$Mo
\cite{expSiMo}.

The comparison between the theoretical values and the experimental
data for the mean angular momentum $<L(E)>=\sum_\ell \ell
\sigma(E,\ell)/ \sigma(E)$ for the reactions
$^{28}$Si+$^{94,100}$Mo is also presented in Fig. 4. The
calculation taking into account both the low-energy surface
vibrational $2^+$ and $3^-$ states and the transfer channels
describes the experimental data for $<L(E)>$ for these reactions.
The coupling to the transfer channels enhances $<L(E)>$ near the
barrier and leads to a bump in $<L(E)>$ at energies below barrier.

The reactions $^{28}$Si+$^{94,100}$Mo are also analyzed in
\cite{expSiMo} by taking into account the coupling to the $2^+$
and $3^-$ surface vibrational states and to the 2-neutron transfer
channel. The 2-neutron transfer channel is treated in
\cite{expSiMo} phenomenologically. The theoretical curves obtained
in \cite{expSiMo} also describe well the experimental data for the
$^{28}$Si+$^{94,100}$Mo reactions.

The comparison of the theoretical curves with the experimental
data in Figs. 1,3,4 shows that our model describes the entrance
channel effects in the subbarrier fusion of a nuclei located along
the $\beta$-stability line. The neutron transfer channels are very
important for the reactions $^{40}$Ca+$^{96}$Zr and
$^{28}$Si+$^{94,100}$Mo near and especially below barrier, see
Figs. 3-4.

\section{Subbarrier fusion of nuclei far from the $\beta$-stability
line}

The subbarrier fusion reactions between a nucleus near to the
proton drip line and a nucleus near to the neutron drip line
should be the most strongly enhanced by the few-nucleon transfer
because in this case the separation energies of the transferred
particles are small and the $Q$-values of the transfer reactions
have very large positive values. However, it is difficult to
perform experiments for such systems. Therefore we study the
fusion reaction between a nucleus near to the neutron drip line
and a $\beta$-stable nucleus in this section. The fusion cross
section for such systems may be measured by using radioactive ion
beams.

The fusion cross sections for the reactions
$^{16,18,20,22,24}$O+$^{58}$Ni obtained in our model are shown in
Figs. 5 and 6. The parameters of the low-energy $2^+$ and $3^-$
states for $^{16,18}$O and $^{58}$Ni taken from \cite{lowstate}
are listed in Table 1. The experimental energies of the first
$2^+$ and $3^-$ states for the neutron-rich isotope $^{20}$O are
known \cite{lowstate} but the experimental deformations $\beta_2$
and $\beta_3$ of these states are not available. Therefore in the
calculation we take the same deformations $\beta_2$ and $\beta_3$
for $^{20}$O as for $^{18}$O. The experimental data for both
energies and the deformation parameters of $2^+$ and $3^-$ states
for extremely neutron-rich isotopes $^{22,24}$O are absent. Due to
this we take the same values of these parameters as for $^{20}$O
in calculations.

We determine the values of $r_0$ in the KNS potential (see Table
1) and take $\delta =1.2$ fm for all reactions with oxygen by
using the experimental fusion cross section \cite{expONi} near
barrier for the reactions $^{16,18}$O+$^{58}$Ni. (Note that we
have taken $\delta =0.7$ fm for all reactions discussed above.)
The wave functions of the neutrons above the magic number 8 in
oxygen isotopes are located in the surface region. Therefore due
to the small separation energies of the neutrons in the
neutron-rich oxygen isotopes and due to the surface localization
of the transferred neutrons the large value of $\delta$ is
reasonable for these reactions. The values of $r_0$ for the oxygen
isotopes in Table 1 are slightly smaller then recommended in
\cite{KNS}. Note that similar values of $r_0$ are also found in
the phenomenological analysis of these reactions in \cite{expONi}.
The experimental data for $^{16,18}$O+$^{58}$Ni are well described
in the framework of our model, see Fig. 5.

The fusion cross sections for $^{16,18,20,22,24}$O+$^{58}$Ni
slightly increase with the number of neutrons for energies near the
barrier. The fusion cross sections below barrier are strongly
enhanced by the few-neutron transfer from oxygen to nickel. We
take into account channels with 1-, 2-, 3- and 4-neutrons transfer
in calculations for the reactions $^{18,20,22}$O+$^{58}$Ni. For
the reaction $^{24}$O+$^{58}$Ni we employ only three transfer
channels related to 1-, 2- and 3-neutron transfer. The influence
of few-neutron transfer channels is important below barrier, see
Figs 5-6. The enhancement of subbarrier fusion cross section due
to neutron transfer channel increases with the number of neutrons
in oxygen.

The partial contributions of the channels with 1-, 2- and
3-neutron transfer to the total fusion cross sections
$^{24}$O+$^{58}$Ni are shown in Fig. 6. We may conclude that the
1-neutron transfer channel is important for energies near the
barrier, the 2-neutron transfer channel gives dominant
contributions for low energies and the 3-neutron transfer channel
give small contributions for energies larger than 20 MeV.

The energy dependence of the mean angular momentum $<L(E)>$ of the
compound nucleus formed in the fusion reaction $^{24}$O+$^{58}$Ni
in different approaches is shown in Fig. 6. We can see in Fig. 6
that the 1-neutron transfer channel enhances $<L(E)>$ near barrier
and the 2-neutron transfer channel leads to the maximum in
$<L(E)>$ at subbarrier energies.

Note that the fusion cross sections for systems $^{24}$O+$^{58}$Ni
and $^{40}$Ca+$^{96}$Zr have different behaviors near barrier due
to the 1-neutron transfer channel contribution. This channel is
not important for the reaction $^{40}$Ca+$^{96}$Zr in contrast to
the reaction $^{24}$O+$^{58}$Ni. Such difference is related to the
$Q$-values of the 1-neutron transfer channel for these reactions:
$Q_{1n}=5.29$ MeV for $^{24}$O+$^{58}$Ni and $Q_{1n}=0.508$ MeV for
$^{40}$Ca+$^{96}$Zr.

The energy dependence of the fusion cross sections and of the mean
angular momentum for the reactions
$^{28}$Si+$^{124,126,128,130,132}$Sn are presented in Fig. 7. In
the calculations we take into account both the 1-, 2-, 3- and
4-neutron transfer channels from the tin isotopes to silicon and
the coupling to $2^+$ and $3^-$ low-energy states. We see in Fig.
7 that $\sigma_{\rm fus}(E)$ and $<L(E)>$ increase drastically
with the number of neutrons in the tin isotopes due to the
two-neutron transfer below barrier.

Our calculation of $\sigma_{\rm fus}(E)$ and $<L(E)>$ for the
reactions $^{28}$Si+$^{124,126,128,130,132}$Sn is done for
$\delta=0.7$ fm, because the $^{132}$Sn is a double magic nucleus.
The values of the parameters of the $2^+$ and $3^-$ low-energy
states and the radii $r_0$ in the KNS potential are listed in
Table 1. Note that in the case of tin isotopes we know the values
of the energies and the deformation parameters $\beta_{2,3}$ of
the $2^+$ and $3^-$ states for $^{124}$Sn only, see
\cite{lowstate}. The experimental deformations $\beta_{2,3}$ for
the isotopes $^{126,128,130,132}$Sn are not known \cite{lowstate}.
Therefore we take the same values of $\beta_2$ and $\beta_3$ for
these isotopes as for $^{124}$Sn. The energies of the $2^+$ and
$3^-$ states in $^{124,126,128,130}$Sn are located around 1.1-1.2
MeV and 2.5-2.8 MeV, respectively \cite{lowstate}. The known
experimental energies of the first $2^+$ and $3^-$ states in
$^{132}$Sn \cite{lowstate} are higher than the corresponding
ranges of energies of the first $2^+$ and $3^-$ states in
$^{124,126,128,130}$Sn. For the reason of systematic of the
excitation energies in the tin isotopes we take the same energies
of the $2^+$ and $3^-$ states for $^{132}$Sn as for $^{130}$Sn.

\section{Conclusions}

We have analyzed the subbarrier fusion cross sections, the mean
and angular momentum induced by heavy ions collisions. The
coupling to the low-energy surface vibrational states and to the
subbarrier few-neutron transfer channels are taken into account in
our model.

It is shown that the few-nucleon transfer with a large positive
$Q$-value leads to strong enhancement of $\sigma_{\rm fus}(E)$ and
$<L(E)>$. Due to few-nucleon transfer the slope of the fusion
cross section changes and a non-monotonous energy dependence
appears in $<L(E)>$.

The few-nucleon transfer enhancement of subbarrier fusion
reactions is very important for the case of systems with small
separation energies of the transferred particles and with large
positive $Q$-value of the transfer reactions. The favorable
conditions for this enhancement take place in reactions between a
nucleus near the neutron drip line and a nucleus along the line of
$\beta$-stability. As a rule, the most important contributions
to the few-nucleon transfer enhancement of the subbarrier fusion
are related to the 1- and 2-nucleon transfer channels.

Our model has been applied to the few-neutron transfer case in
this paper. The inclusion of few-proton transfer in our model is
direct because it is necessary to take into account the Coulomb
interactions between the ions and the transferred protons in the
action (13). An extension of our model to the case of
neutron-proton transfer is also straightforward but the sequence
of the neutron-proton transfer should be taken into account due to
the different values of the separation energies of nucleons for
different sequences of neutron-proton transfer.

Our calculation is based on the parameterization (18) of the
ion-ion interaction at small distance. The parameterization
(18) is matched with the Coulomb and the KNS nuclear interactions
in Sec. 2. Note that this parameterization may be matched with
any other potentials and may be used for the description of other
reactions near the barrier.

We consider sequential transfer of neutrons. This transfer
mechanism is may apply for colliding nuclei located not very far
from $\beta$-stability line. For extremely far from the
$\beta$-stability line colliding nuclei probably is necessary to
take into account correlated transfer of neutrons \cite{reisdorf}.
The correlated neutron transfer may be especially important in the
case of fusing system with two-neutron halo nucleus. It is
possible to consider correlated transfer of neutrons in the
framework of the model also if we change the action (13)
correspondingly.

\begin{acknowledgments}
The author would like to thank R.W. Hasse and F.A. Ivanyuk for
careful reading of manuscript. He acknowledges gratefully support
from GSI. He also would likes to thank A.M. Stefanini for
bringing the Table of experimental data measured in \cite{expCaZr}
to his attention before publication. \end{acknowledgments}

\begin{figure}
\caption{ Fusion cross sections for the reactions
$^{28}$Si+$^{58,62,64}$Ni (top) and $^{30}$Si+$^{58,62,64}$Ni
(bottom). Experimental data (dots) are taken from [29]. The
results of the calculation taking into account both the low-energy
$2^+$ and $3^-$ states and the neutrons transfer channels are
shown by the solid curves. The results of the calculation
taking into account the coupling to the low-energy $2^+$ and $3^-$
states are marked by the dash curves. The dash-dots curves
correspond to the calculation with transfer channels only and the
results of the calculation in the one-dimensional WKB approach are
shown by dotted curves. } \end{figure}

\begin{figure}
\caption{ Effective potential for the reaction
$^{28}$Si+$^{94,100}$Mo for the case $\ell=0$ and $\epsilon_k=0$
without and with 1, 2, 3 and 4 neutrons transfer from
$^{100}$Mo to $^{28}$Si and the effective potential for reactions
$^{28}$S+$^{94}$Mo without neutron transfer for $\ell=0$ and
$\epsilon_k=0$.} \end{figure}

\begin{figure}
\caption{Fusion cross sections for the reactions
$^{40}$Ca+$^{96}$Zr (top) and $^{40}$Ca+$^{90}$Zr (bottom).
Experimental data (dots) are taken from [20]. The notations are
the same as in Fig. 1. } \end{figure}

\begin{figure}
\caption{Fusion cross section (top) and mean angular momentum
(bottom) for the reactions $^{28}$Si+$^{100}$Mo (left) and
$^{28}$Si+$^{94}$Mo (right). Experimental data (dots) are taken
from [30]. The notations are the same as in Fig. 1.}
\end{figure}

\begin{figure}
\caption{Fusion cross sections for the reactions
$^{16,18,20,22}$O+$^{58}$Ni. Experimental data (dots) for
reactions $^{16,18}$O+$^{58}$Ni are taken from [33].
The notations are the same as in Fig. 1. } \end{figure}

\begin{figure}
\caption{Fusion cross section (top) and mean angular momentum
(bottom) for reaction $^{24}$O+$^{58}$Ni. The partial
contributions of channels with 1, 2 and 3 neutrons transfer to the
total cross section are marked by squares and ellipses in the
cases with and without contributions related to $2^+$ and $3^-$
low-energy excited states, respectively. The other notations are
the same as in Fig. 1. } \end{figure}

\begin{figure}
\caption{Fusion cross section (top) and mean angular momentum
(bottom) for the reactions $^{28}$Si+$^{124,126,128,130,132}$Sn.}
\end{figure}

\begin{table}
\caption{ Excitation energies $E_{\ell}$, deformation parameters
$\beta_\ell$ and multipolarities $\ell$ of the low-energy surface
vibrational states and the values of the parameter $r_0$ in
the $V_{\rm KNS}(r)$ nuclear potential [41]. }

\begin{tabular}{ccccc}
Nucleus & $E_{\ell}$ (MeV) & $\beta_\ell$& $\ell$ & $r_0$ (fm) \\
\hline
$^{16}$O & 6.92 & 0.36 & 2 & 1.11 \\
$^{16}$O & 6.13 & 0.60 & 3 & 1.11 \\
$^{18}$O & 1.98 & 0.39 & 2 & 1.11 \\
$^{18}$O & 5.10 & 0.48 & 3 & 1.11 \\
$^{20}$O & 1.63 & 0.39 & 2 & 1.11 \\
$^{20}$O & 5.62 & 0.48 & 3 & 1.11 \\
$^{28}$Si & 1.78 & 0.41 & 2 & 1.165 \\
$^{28}$Si & 6.88 & 0.39 & 3 & 1.165 \\
$^{30}$Si & 2.24 & 0.22 & 2 & 1.195 \\
$^{30}$Si & 5.59 & 0.15 & 3 & 1.195 \\
$^{40}$Ca& 3.90 & 0.11 & 2 & 1.21 \\
$^{40}$Ca& 3.74 & 0.34 & 3 & 1.21 \\
$^{58}$Ni & 1.45 & 0.18 & 2 & 1.16 \\
$^{58}$Ni & 4.47 & 0.22 & 3 & 1.16 \\
$^{62}$Ni & 1.17 & 0.22 & 2 & 1.18 \\
$^{62}$Ni & 3.76 & 0.14 & 3 & 1.18 \\
$^{64}$Ni & 1.34 & 0.17 & 2 & 1.19 \\
$^{64}$Ni & 3.56 & 0.15 & 3 & 1.19 \\
$^{90}$Zr& 2.19 & 0.08 & 2 & 1.21 \\
$^{90}$Zr& 2.75 & 0.14 & 3 & 1.21 \\
$^{96}$Zr& 1.76 & 0.12 & 2 & 1.22 \\
$^{96}$Zr& 1.91 & 0.22 & 3 & 1.22 \\
$^{94}$Mo & 0.87 & 0.128 & 2 & 1.17 \\
$^{94}$Mo & 2.53 & 0.161 & 3 & 1.17 \\
$^{100}$Mo& 0.534 & 0.226 & 2 & 1.18 \\
$^{100}$Mo& 1.91 & 0.21 & 3 & 1.18 \\
$^{124}$Sn& 1.13 & 0.076 & 2 & 1.18 \\
$^{124}$Sn& 2.59 & 0.072 & 3 & 1.18 \\
$^{126}$Sn& 1.14 & 0.076 & 2 & 1.18 \\
$^{126}$Sn& 2.72 & 0.072 & 3 & 1.18 \\
$^{128}$Sn& 1.17 & 0.076 & 2 & 1.18 \\
$^{128}$Sn& 2.76 & 0.072 & 3 & 1.18 \\
$^{130}$Sn& 1.22 & 0.076 & 2 & 1.18 \\
$^{130}$Sn& 2.49 & 0.072 & 3 & 1.18 \\
$^{132}$Sn& 1.22 & 0.076 & 2 & 1.18 \\
$^{132}$Sn& 2.49 & 0.072 & 3 & 1.18 \\
\end{tabular}

\end{table}

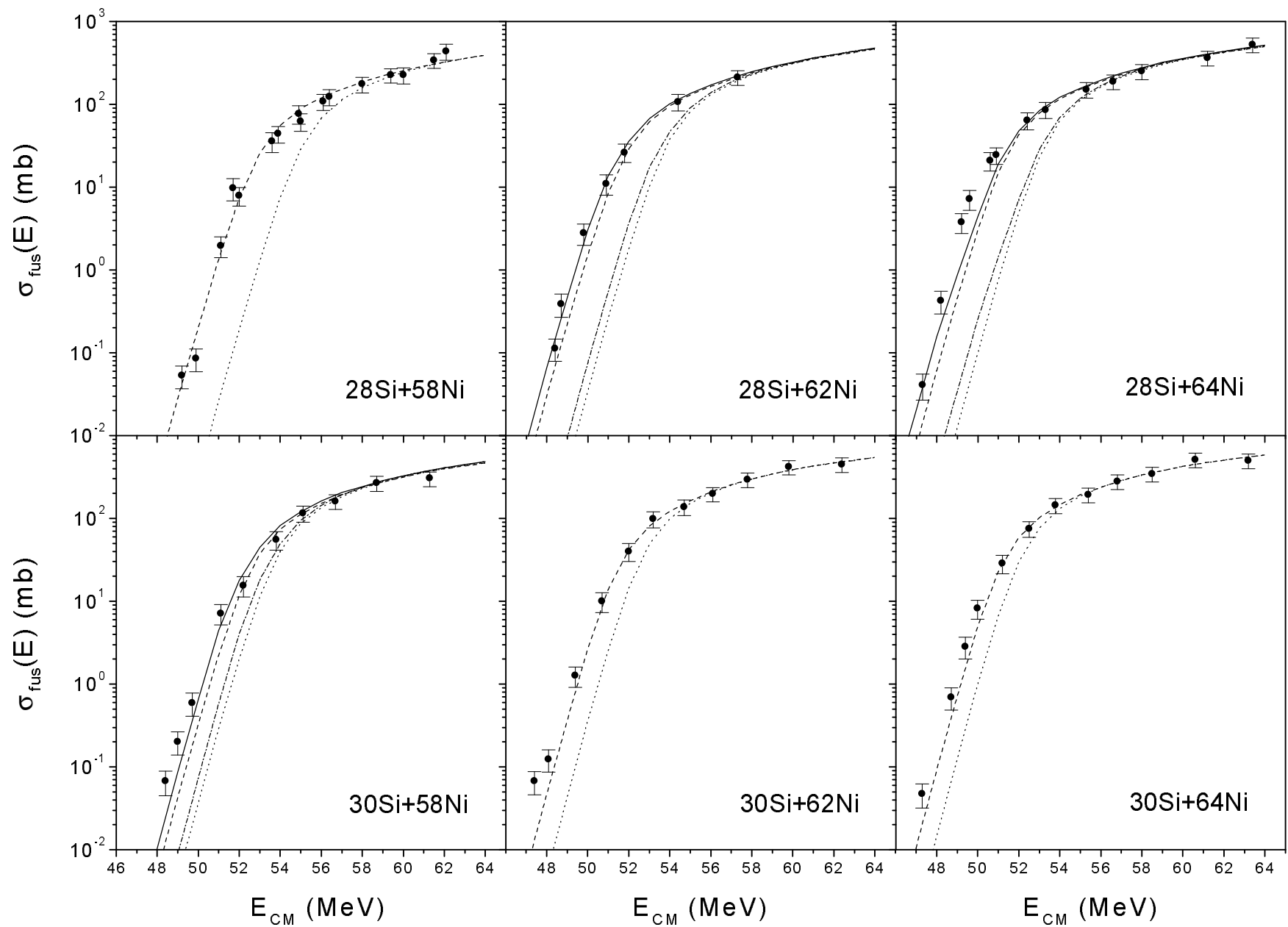
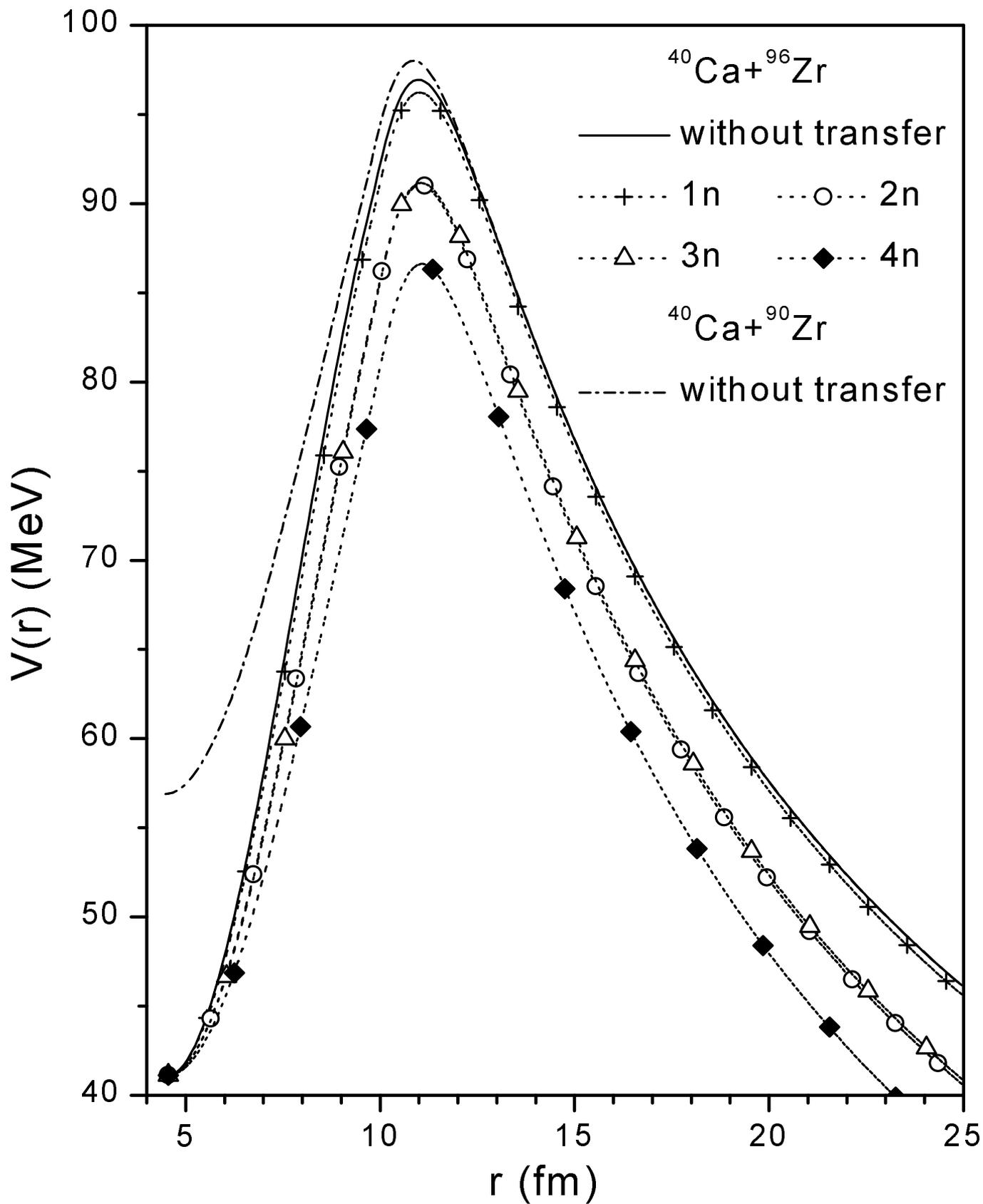
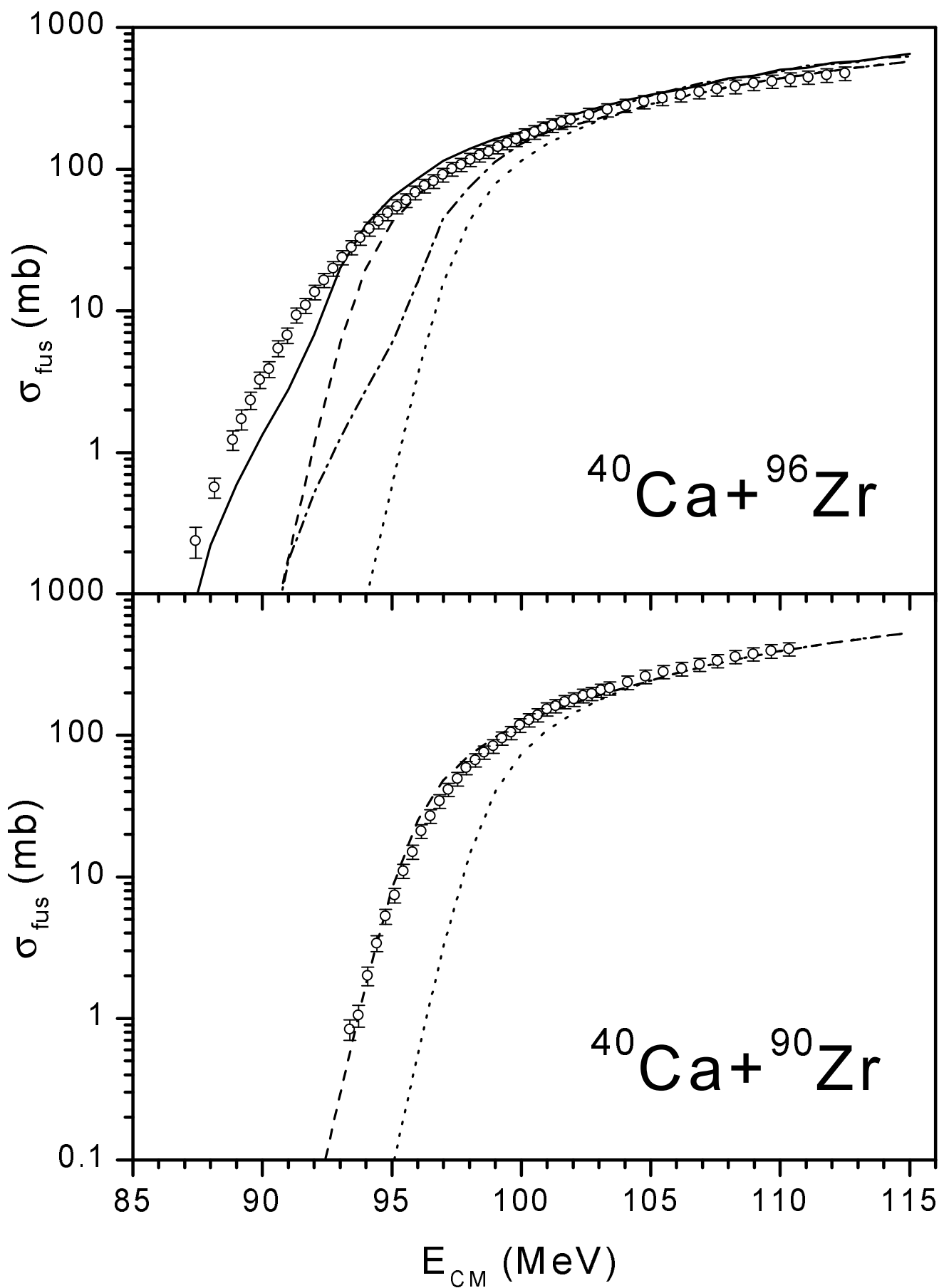
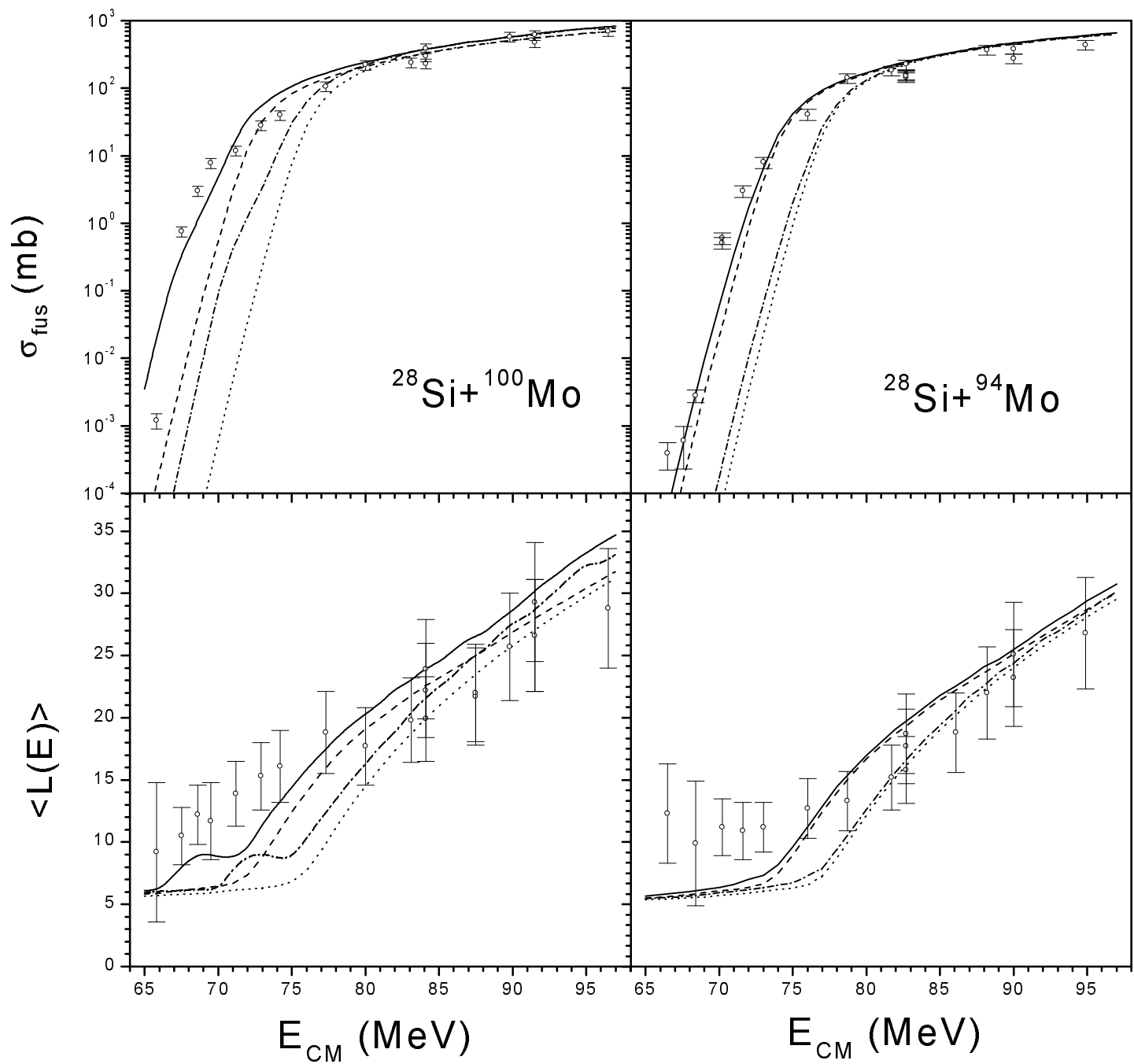
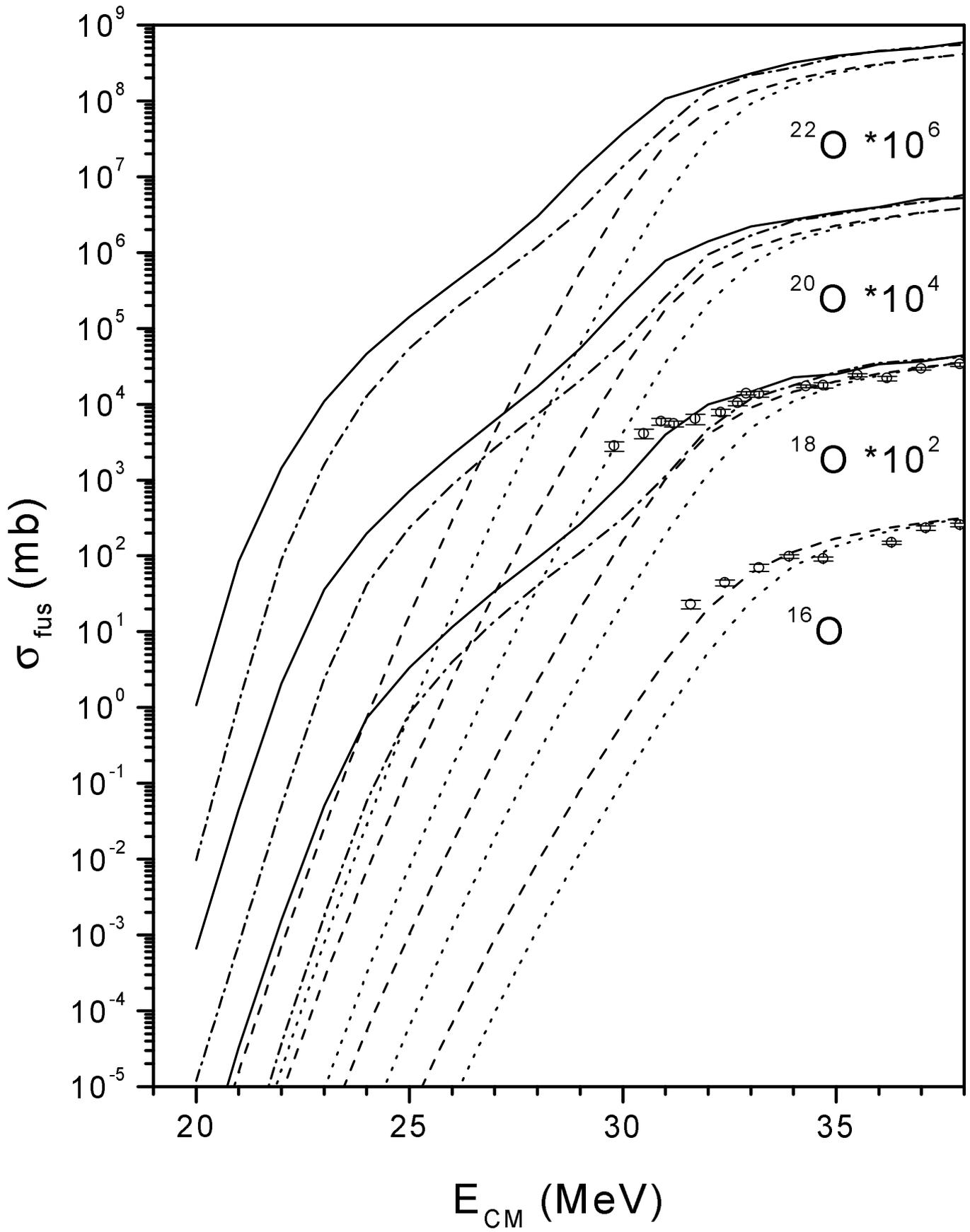
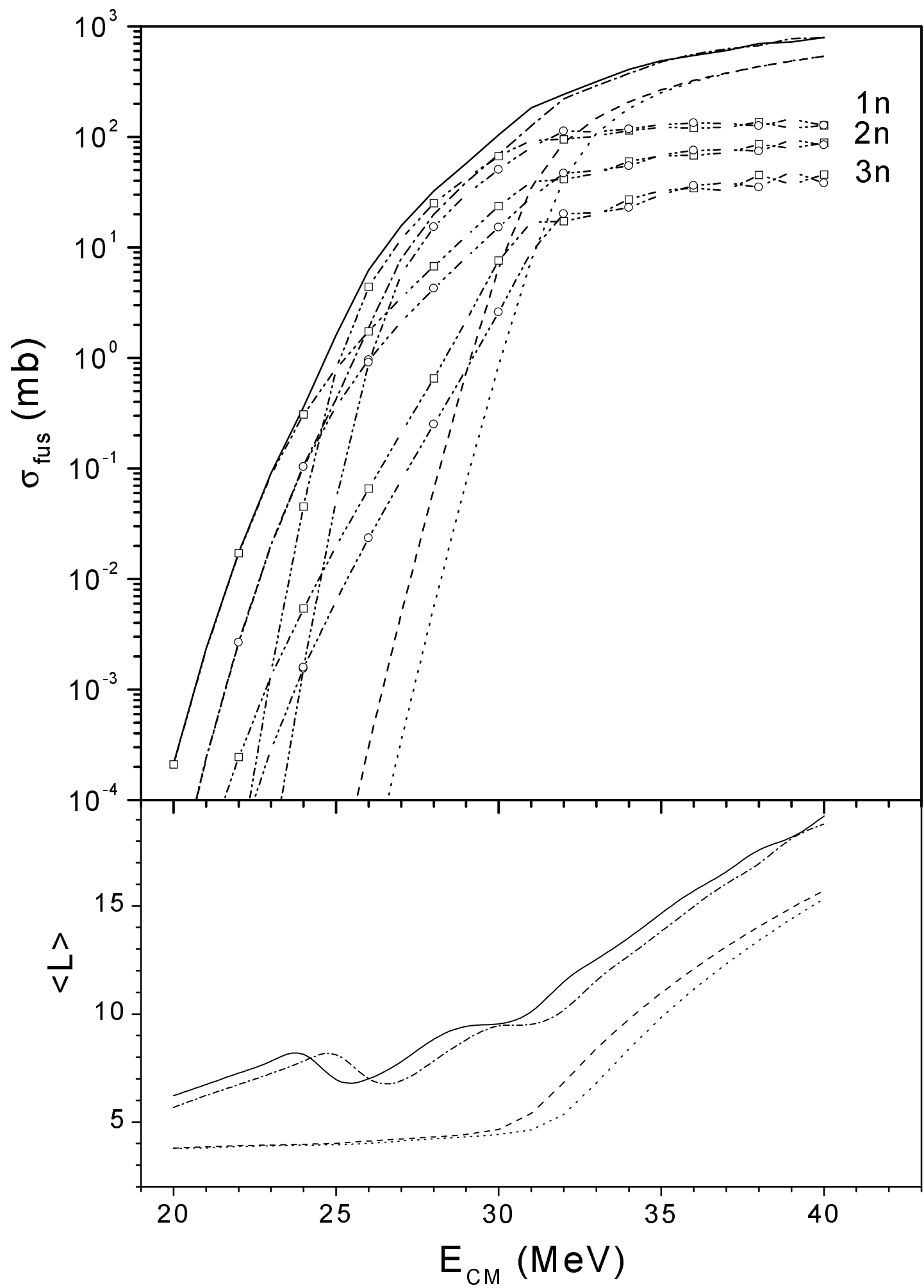
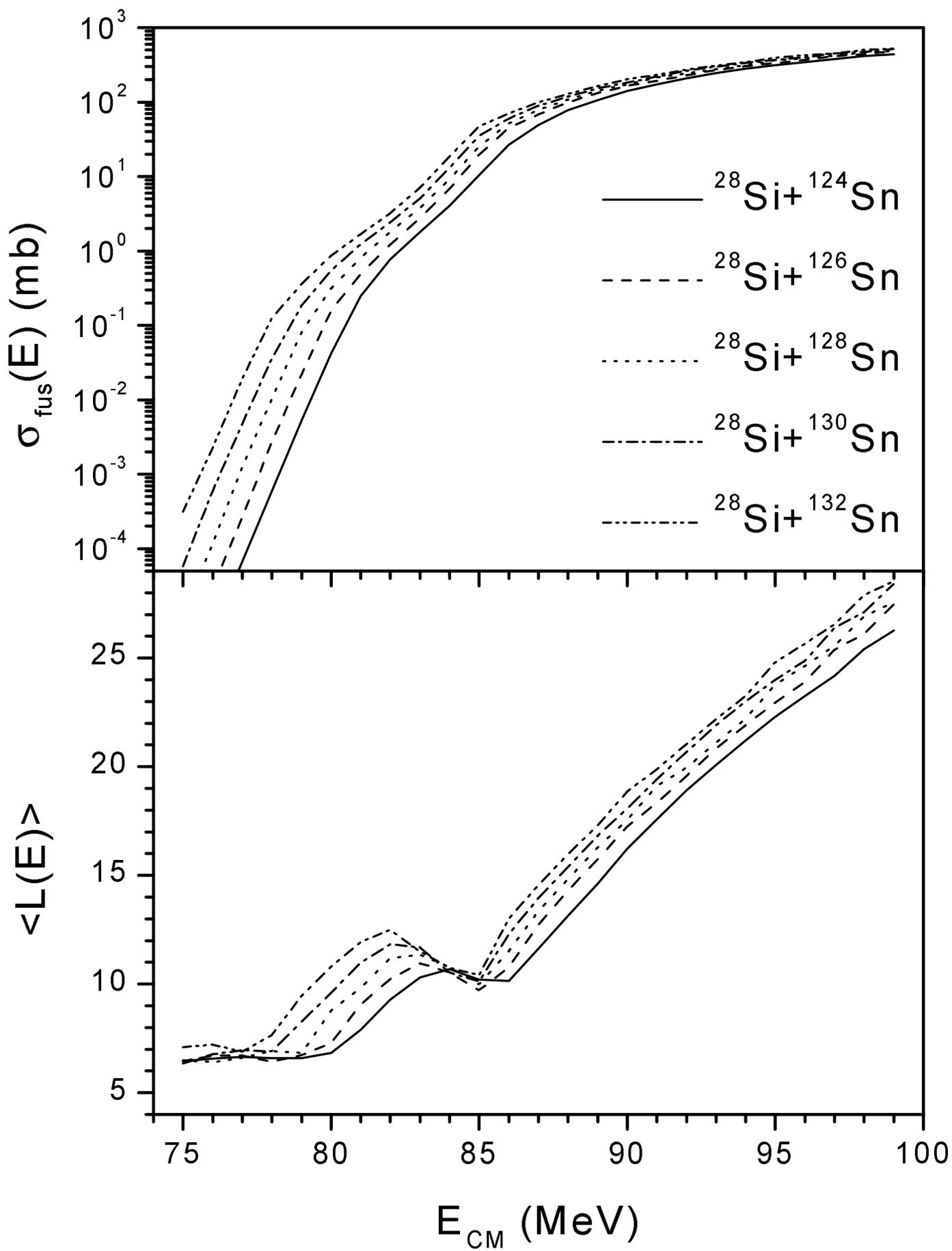
\end{document}